\documentclass[aps,prl,reprint]{revtex4-1}
\usepackage[utf8]{inputenc}
\usepackage{natbib}
\usepackage{graphicx}
\usepackage{xcolor}
\usepackage{bm}
\usepackage{amsmath}
\usepackage{amssymb}
\usepackage{subfig}
\usepackage{float}
\newcommand{\vect}[1]{\bm{#1}}
\newcommand{\ten}[1]{\mbox{\textbf{\textit{\textsf{#1}}}}}

\newcommand{\tenszero}{\mbox{\textbf{\textsf{0}}}}
\newcommand{\sprod}{\!\cdot\!}

\newcommand{\vprod}{\!\times\!}

\newcommand{\dif}{\mathrm{d}}
\newcommand{\mi}{\mathrm{i}}
\newcommand{\me}{\mathrm{e}}
\newcommand{\GRe}{\mathcal{R}\mathrm{e}}
\newcommand{\GIm}{\mathcal{I}\mathrm{m}}
\begin{document}

\title{Quantum sensing protocol for motionally chiral Rydberg atoms}
\author{Stefan Yoshi Buhmann$^1$}
\email{stefan.buhmann@physik.uni-freiburg.de}
\author{Steffen Giesen$^2$}
\author{Mira Diekmann$^2$}
\author{Robert Berger$^2$}
\author{Stefan Aull$^3$}
\author{Markus Debatin$^3$}
\author{Peter Zahariev$^{3,4}$}
\author{Kilian Singer$^3$}
\email{ks@uni-kassel.de}
\affiliation{$^1$Theoretische Physik III, Universit\"at Kassel, Heinrich-Plett-Str. 40, 34132 Kassel, Germany}
\address{$^2$Fachbereich Chemie, Philipps-Universit\"at Marburg, Hans-Meerwein-Str 4, Marburg 35032, Germany}
\address{$^3$Experimentalphysik I, Universit\"at Kassel, Heinrich-Plett-Str. 40, 34132 Kassel, Germany}
\address{$^4$Institute of Solid State Physics, Bulgarian Academy of Sciences,  72, Tzarigradsko Chaussee, 1784 Sofia, Bulgaria}

\begin{abstract}
A quantum sensing protocol is proposed for demonstrating the motion-induced chirality of circularly polarised Rydberg atoms. To this end, a cloud of Rydberg atoms is dressed by a bichromatic light field. This allows to exploit the long-lived ground states for implementing a Ramsey interferometer in conjunction with a spin echo pulse sequence for refocussing achiral interactions. Optimal parameters for the dressing lasers are identified. Combining a circularly polarised dipole transition in the Rydberg atom with atomic centre-of-mass motion, the system becomes chiral. The resulting discriminatory chiral energy shifts induced by a chiral mirror are estimated using a macroscopic quantum electrodynamics approach.
\end{abstract}

\maketitle

\date{August 2020}

\paragraph{Introduction.}
We propose a method for inducing and detecting chirality in Rydberg atoms by combining circular dipole transitions with centre-of-mass motion. The scheme is based on measuring the dispersion interaction of this artificial chiral system with a chiral mirror. The predicted discriminatory interaction is a leading-order relativistic quantum electrodynamics effect.

\begin{figure}[t]
\centering
\includegraphics[width=0.35\textwidth]{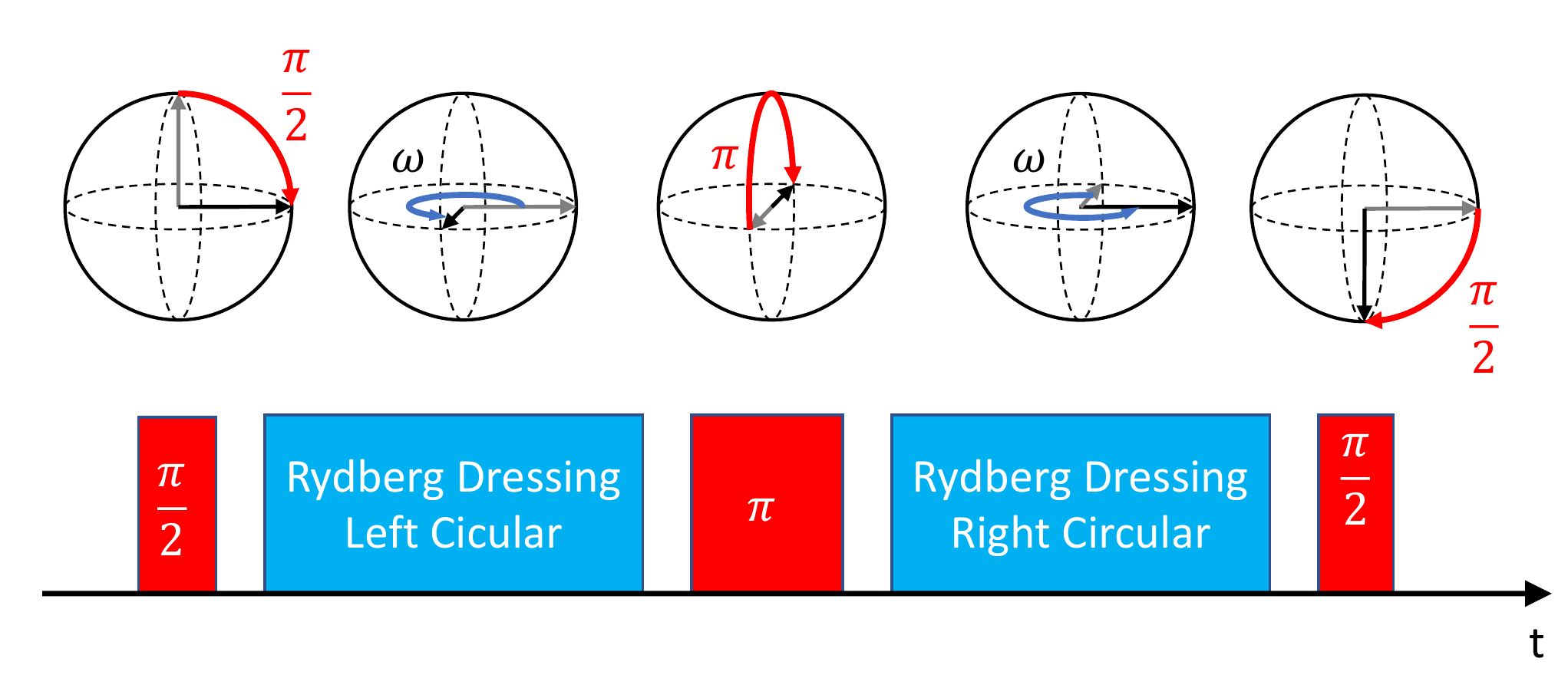}
\caption{Quantum-sensing protocol for detecting chirality: Manipulation of the electronic ground-state hyperfine manifold of ${}^{87}$Rb by microwaves (red) of 6.8~GHz, dressing lasers (blue) are switched in circularity after a central spin echo pulse (red). A continuous chiral influence reveals the motional chirality of the Rydberg atoms.}
\label{fig:ramseyb}
\end{figure}

Rydberg atoms in conjunction with dressing represent a formidable quantum information platform \cite{Jau2016,PhysRevA.101.030301} and quantum sensor due to interaction-induced energy shifts caused by externally applied fields \cite{PhysRevLett.122.053601} and nearby particles such as molecules. We suggest to use dressed Rydberg atoms as sensitive quantum sensors \cite{Adams_2019} for chirality exploiting the long lifetimes of sensitive Rydberg states and accurate clock states featured in the hyperfine manifold of the electronic ground state of alkali metals, which are the basis of the current definition of time.
Using two-photon dressing from one of the hyperfine levels of the electronic ground states to a Rydberg state we can combine an accurate reference with a sensitive Rydberg state such that interaction-induced energy shifts \cite{doi:10.1063/1.1928850,doi:10.1142/9789812701473_0027,doi:10.1002/3527603417.ch15} are not probed directly by spectroscopy on the Rydberg state \cite{PhysRevLett.93.163001,Singer_2005a,Singer_2005b} but are translated to energy shifts of the clock states. Employing a Ramsey sensing sequence \cite{ramsey:1950} these energy shifts can be accurately determined (Fig.~\ref{fig:ramseyb}).

Dispersive energy shift in atoms may arise from their interaction with polarisable objects such as molecules \cite{London30,SYBRef5} or surfaces \cite{SYBRef9,SYBRef5}. Linear-response or macroscopic quantum electrodynamics frameworks allow to consider objects of arbitrary shapes and materials
\cite{Agarwal75,Wylie85,SYBRef2}. A range of different materials has been considered both for Casimir--Polder and the closely related Casimir interaction \cite{Woods16}, including magnetoelectric metamaterials \cite{Kenneth02,Henkel05,Rosa08}, chiral objects \cite{Jenkins94,Craig1999,Butcher12,Barcellona17}, or topological insulators \cite{Grushin11,Fuchs17b,Fuchs17}. Rydberg atoms with their large electric transition dipole moments \cite{Gallagher05,yerokhin2016} are particularly susceptible to dispersive energy shifts. For this reason, they where instrumental in both the first experimental verification of the retarded Casimir--Polder interaction of atoms \cite{Sukenik93} and the later demonstration of thermal effects in this interaction \cite{Marrocco98}. On the theory side, it has been shown that the long wavelength of photons arising from neighbouring Rydberg-level transitions can render Casimir--Polder forces good conductors almost independently of temperature \cite{Ellingsen10} and that their extreme size leads to multipole contributions particularly at small separations \cite{SYBRef11}. Recent works have proposed sensitive probing of Rydberg--surface interactions via electromagnetically induced transparency \cite{Yang19}, predicted that surface-induced Casimir--Polder interactions can lead to the formation of Rydberg dimers \cite{Block19}, and even suggested that Rydberg atoms could be used as sensors for the dynamical Casimir effect \cite{Antezza14}.

To induce chirality in an achiral Rydberg atom, we combine a circularly polarised dipole transition of the Rydberg atom with atomic centre-of-mass motion. When the time-reversal odd axial vector representing the circular dipole transition has a nonzero component along the time-reversal-odd polar centre-of-mass velocity, the combined system becomes time-reversal even and parity odd, thus realizing true chirality according to Barron \cite{Barron2013}. Attempts to utilise such motionally chiral Rydberg atoms to detect an asymmetry in the charge transfer reaction to chiral molecules have been reported, but remained unsuccessful \cite{hammer:2002}.
Herein instead, we propose to detect the resulting enantiodiscriminatory dispersion interaction due to fluctuating fields of a nearby mirror, which is analogous to predicted enantiodiscriminatory optical forces induced by real fields \cite{canaguier2013,cameron2014}.

\begin{figure}[H]
\centering
\includegraphics[width=0.3\textwidth]{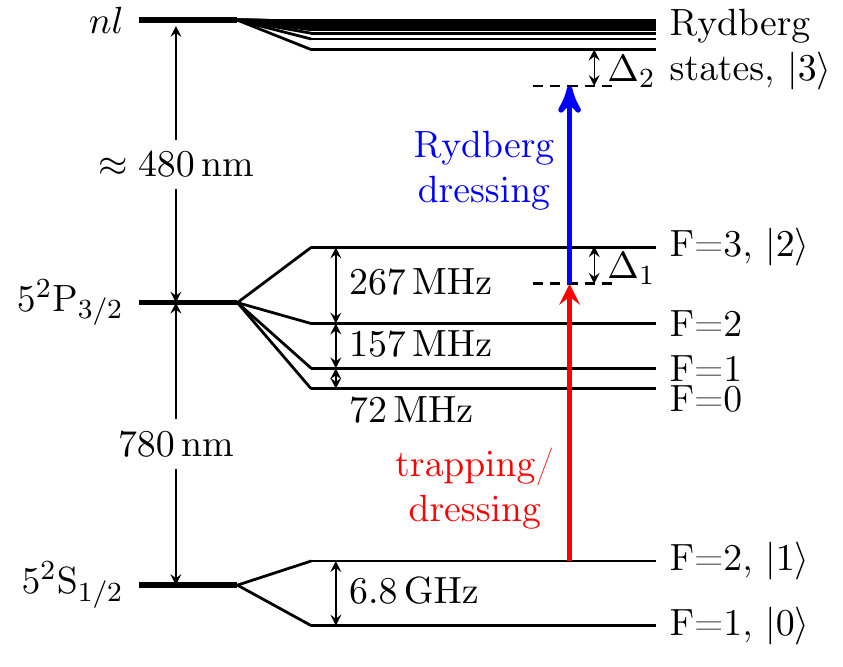}
\caption{Level scheme of 87-Rubidium (not to scale). $\left|0\right>$ is the $5^2S_{1/2},\, F=1$ ground state serving as reference, $\left|1\right>$ is the $5^2S_{1/2},\, F=2$ state which is dressed by a two colour two photon excitation scheme including the intermediate state  $\left|2\right>$ and the Rydberg state $\left|3\right>$. Laser intensities will be given in terms of one photon resonant Rabi-frequencies $\Omega_{12}$ and $\Omega_{23}$ for the 780~nm laser and 480~nm laser respectively.}
\label{fig:ramseya}
\end{figure}

\paragraph{Quantum sensing protocol.}
In the following paragraphs, we will introduce a protocol for sensing energy shifts of Rydberg states using a Ramsey-sequence type of measurement, in which we assume that state-changing collision rates are less than 1~Hz. As can be seen in Fig.~\ref{fig:ramseyb}, the measurement Ramsey sequence is initiated by a $\pi/2$ pulse on the electronic ground state hyperfine manifold of ${}^{87}$Rb which brings the state in an superposition of $5^2S_{1/2},\, F=1$ and $5^2S_{1/2},\, F=2$ states (see Fig.~\ref{fig:ramseya} for a sketch of the relevant level scheme). A subsequent two-photon dressing laser with circularly polarized light is coupling the interaction induced energy shifts caused by chiral systems to the upper ground state ($5^2S_{1/2},\, F=2$). Detuning and Rabi-frequency of the dressing lasers is chosen to optimize the energy shift and coherence properties (see below). Subsequently, a spin echo is implemented by a $\pi$ pulse resonant to the ground state is used to refocus all interactions caused by achiral influences from the environment, such that only enantiodiscriminatory signals are detected. For the signal caused by the interaction with chiral systems to persist, the circularity of the dressing lasers is switched. The final $\pi/2$ pulse on the ground state is transferring the phase shifts caused by the chiral molecule into a measurable signal. By scanning the duration of the exposure of the atom with the dressing lasers, Ramsey fringes will be observed. The period of these fringes is a direct measure for the energy shift.

\begin{center}
\begin{figure}[H]
\includegraphics[width=1.0\textwidth]{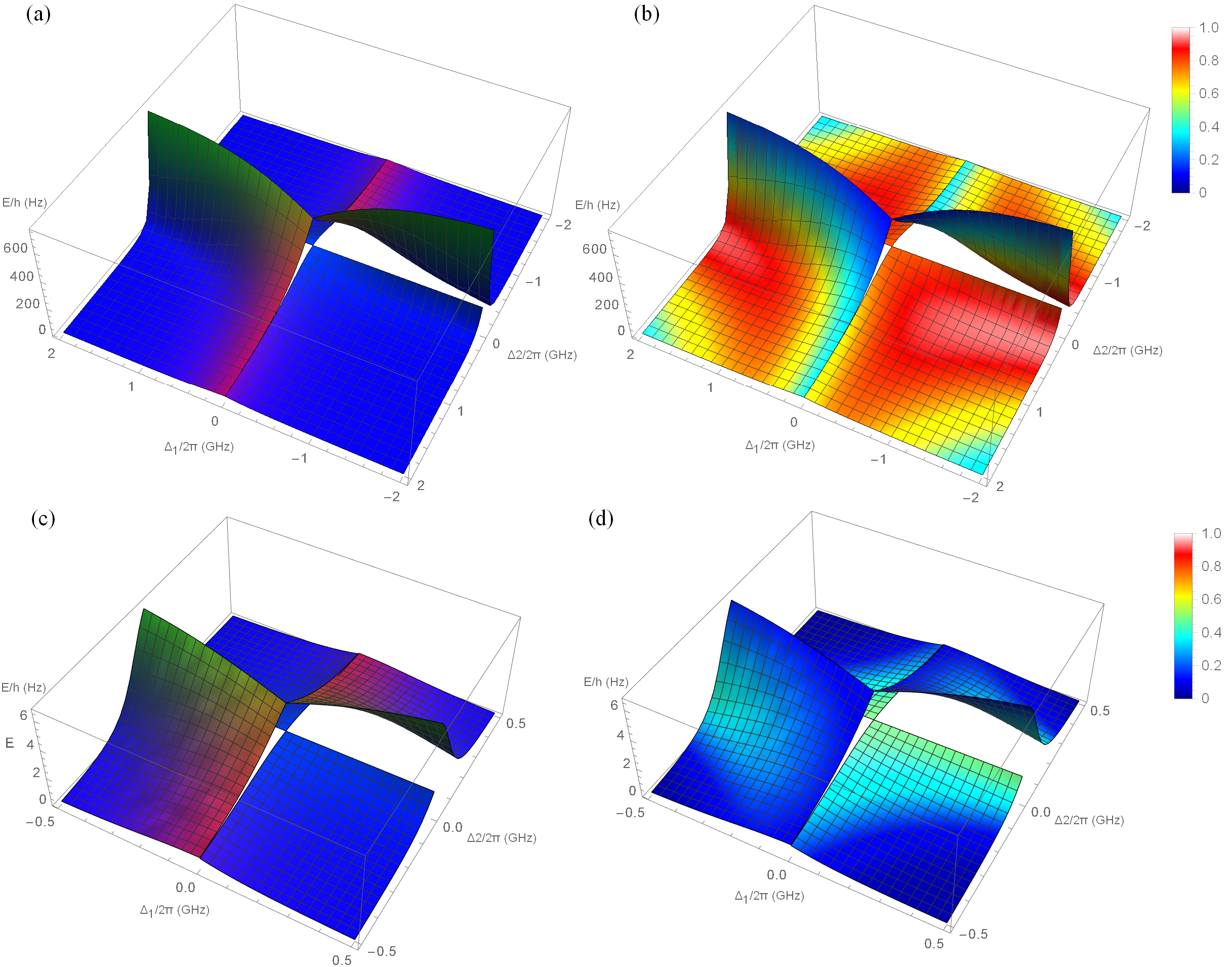}
\caption{Energy shifts of the upper hyperfine component $\left|1\right>$ of the electronic ground state of 87 Rubidium due to two photon dressing with $\Delta_1=(E_{2}-E_{1})/\hbar-\omega_{12}^\mathrm{L}$ being the detuning of the 780~nm laser with respect to the $5^2P_{3/2},\, F=3$ state and $\Delta_2=(E_{3}-E_{1})/\hbar-(\omega_{12}^\mathrm{L}+\omega_{23}^\mathrm{L})$ being the detuning of the two photon transition including the 480~nm laser to the Rydberg state. Positive values of $\Delta$ denote red detuning. First quadrant in each picture with values $\Delta_1$ and $\Delta_2>0$ show the energy shifts of the dressed state with lowest energy, the second quadrant with $\Delta_1<0$ and $\Delta_2>0$ and the fourth quadrant with $\Delta_1>0$ and $\Delta_2<0$ correspond to the intermediate energy dressed eigenstate and the third quadrant with $\Delta_1<0$ and $\Delta_2<0$ depict the energy shifts of the dressed state with highest energy. (a) Energy shift $E$ of the upper ground state due to a $\delta_{\mathrm{RB}}/(2\pi) = 1~\mathrm{kHz}$ shift in the Rydberg state with Rabi frequencies $\Omega_{12}/(2 \pi)=0.5$ GHz and $\Omega_{23}/(2 \pi)=1$ GHz. Blue/red/green colour intensity indicate dominating ground/intermediate/Rydberg state admixture to the corresponding dressed state. (b) Same data is plotted as in (a) but colour is now indicating the amplitude of the second Ramsey peak in relation to first Ramsey peak given by $\sum_i P_i e^{-\Gamma_i / (E/\hbar)}$, with $P_i$ being the probability of being in state $\left| i \right>$. Thus it is better to choose detunings resulting in smaller energy shifts $E$ (white shaded area close to $\Delta_1/(2\pi) = 1.5$GHz and $\Delta_2/(2\pi)=0.2$GHz. (c) Plot description is as in (a) but for a $\delta_{\mathrm{RM}}/(2\pi) = 10~\mathrm{Hz}$ shift in the Rydberg state plotted for a smaller detuning range and reduced Rabi frequencies of $\Omega_{12}/(2 \pi)=0.25$ GHz and $\Omega_{23}/(2 \pi)=0.5$ GHz. (d) Colours in plot are as described in (b) but for parameters given in (c). Note that optimal parameters depend on the expected signal strengths.}
\label{fig:dressing}
\end{figure}
\end{center}

Optimal parameters of dressing lasers are obtained by solving for the eigenvectors and eigenvalues of the following matrix representation of an effective Hamilitonian within the quasi-resonant approximation involving the three states $\left|1\right>$, $\left|2\right>$ and $\left|3\right>$ \cite{ShoreBook1990}:
		\begin{equation}
			\label{eq:3_level_matrix}
			\mathcal{H}\left(\Delta_2\!+\!\delta_{\mathrm{RM}}\right)/\hbar=\left(
			\begin{matrix}
				-\mathrm{i} \Gamma_1 & \frac{1}{2}\Omega_{12} & 0\\
				\frac{1}{2}\Omega_{12} & -\mathrm{i} \Gamma_2+\Delta_1 & \frac{1}{2}\Omega_{23}\\
				0 & \frac{1}{2}\Omega_{23} & -\mathrm{i}\Gamma_3+\Delta_2 + \delta_{\mathrm{RM}}
			\end{matrix}
			\right)
		\end{equation}
with Rabi frequencies $\Omega_{12}$ and $\Omega_{23}$ as described in Fig.~\ref{fig:ramseya} and detunings of both dressing lasers $\Delta_1$ and $\Delta_2$ as defined in Fig.~\ref{fig:ramseya}. $\hbar = h/(2\pi)$ denotes here the reduced Planck constant and $\delta_\mathrm{RM}$ the interaction-induced shift of the Rydberg state (possible chiral shifts of the other states are negligible by comparison in view of the large Rydberg dipole moments), with $\Gamma_i$ being the decay rates. We assumed a conservative value of $\Gamma_1/(2 \pi)=1\,\mathrm{Hz}$ for the upper hyperfine component of the electronic ground state due to collisions, $\Gamma_2/(2 \pi)=3.8\times10^7\,\mathrm{Hz}$  and $\Gamma_3/(2 \pi)=1.4\times10^5\,\mathrm{Hz}$ for a typical Rydberg state. If the Rydberg state shifts by $\delta_\mathrm{RM}$, we obtain the signal by comparing $\mathcal{H}\left(\Delta_2+\delta_\mathrm{RM}\right)$ with $\mathcal{H}\left(\Delta_2\right)$, which corresponds to the numerical derivative of $\mathcal{H}\left(\Delta_2\right)$ with respect to $\Delta_2$ for $\delta_\mathrm{RM}$ near zero. As a result, the energy shift of the upper ground state due to chiral interaction-induced shifts in the Rydberg state are obtained by comparing the eigenvalues of $\mathcal{H}\left(\delta_\mathrm{RM}\right)$ to $\mathcal{H}\left(0\right)$. Fig.~\ref{fig:dressing}a/c show that maximal energy shifts are obtained when $\Delta_2$ is small. But this is not necessarily optimal as close to resonance admixtures of the high Rydberg state and the intermediate state lead to a strong reduction of the Ramsey fringe contrast. The relevant figure of merit is the decrease of the amplitude of the Ramsey fringe after a full oscillation period which is given by the interaction induced energy shift of the upper ground state. For Rabi frequencies of $\Omega_{12}/(2 \pi)=0.5\,\mathrm{GHz}$  and $\Omega_{23}/(2 \pi)=1\,\mathrm{GHz}$ and Rydberg energy shifts of $\delta_\mathrm{RM}=1\,\mathrm{kHz}$, one finds optimal regions (see white shaded areas in Fig.~\ref{fig:dressing}b)). Also note that for smaller Rydberg energy shifts of 10 Hz, the dressing light intensities are halved, and different optimal detuning parameters have to be chosen (see greed areas in Fig.~\ref{fig:dressing}d). The accuracy is limited by the  $T_1$ and $T_2^*$ times (longitudinal and effective transversal relaxation times) in the electronic ground-state hyperfine manifold.

\onecolumngrid
\twocolumngrid

\paragraph{Chiral energy shifts in Rydberg atoms.}
Dispersion interactions between two objects can possess chiral components, as stated in the introduction. Curie's symmetry principle \cite{SYBRef1} dictates that for these to be enantiodiscriminatory, both interacting objects need to be chiral. In other words, a Rydberg atom can only acquire a chiral energy shift if itself exhibits handedness. This is typically not the case for Rydberg atoms and the associated dominant electric dipole-electric dipole interactions.

\begin{figure}
\centering
\includegraphics[width=0.3\textwidth]{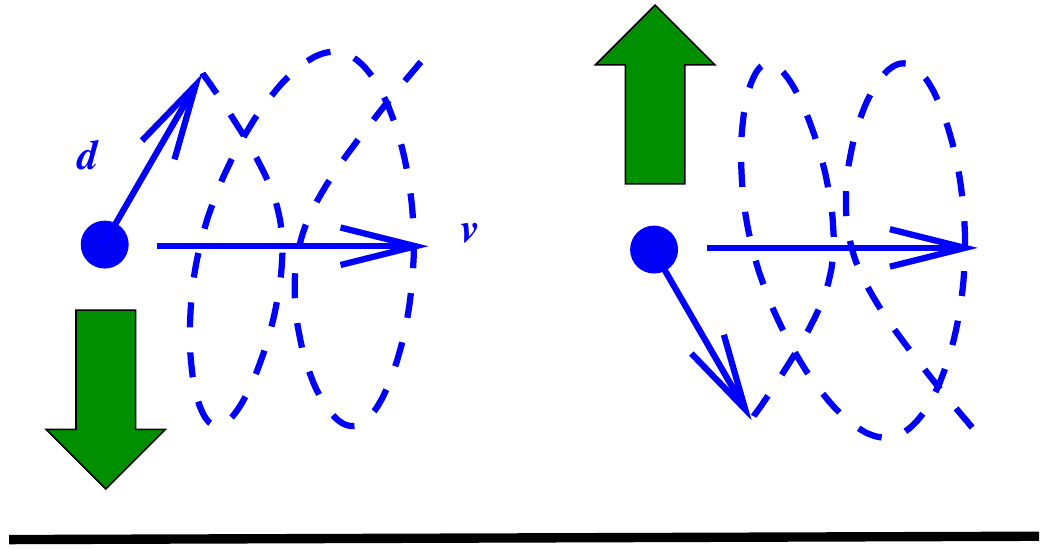}
\caption{Enantiodiscriminatory interactions of chiral Rydberg atoms: the rotating electric dipole moment $\vect{d}$ of a Rydberg atom moving with velocity $\vect{v}$ in the $x$-direction maps out a corkscrew trajectory, forming a chiral system that experiences an enantiodiscriminatory interaction with a perfect chiral mirror with surface normal defining the $z$-axis. Upon reversing the rotation direction of the dipole, the opposite enantiomer is realised, reversing the sign of the interaction.}
\label{fig:SYB1}
\end{figure}

In this section, we present possible a solution to this challenge: a Rydberg atom prepared in a state with a nonvanishing $x$-component $m$ of its orbital angular momentum, with $x$ being arbitrarily chosen here as our quantisation axis, will preferentially or even exclusively undergo circularly polarised electric dipole transitions of a given rotation direction. If the atom moves parallel to the respective rotation axis, then its electric dipole moment maps out a corkscrew trajectory during a transition, with this corkscrew constituting a handed system (Fig.~\ref{fig:SYB1}). This artificial chiral Rydberg system is transformed into its opposite enantiomer upon reversing either the direction of motion or the sign of the $x$-component of orbital angular momentum (and hence the rotation direction of the circular dipole transition). In the following, we will show that this system indeed exhibits discriminatory chiral dispersive energy shifts and estimate the order of magnitude of such shifts as induced by a chiral mirror (Fig.~\ref{fig:SYB1}).

The interaction of a Rydberg atom $\mathrm{A}$ at instantaneous position $\vect{r}_\mathrm{A}$ with velocity $\vect{v}$ with the quantum electromagnetic field $\mathrm{F}$ in electric-dipole approximation is given by \cite{SYBRef2}
$\hat{H}_\mathrm{AF}
=-\hat{\vect{d}}\sprod\hat{\vect{E}}(\vect{r}_\mathrm{A})
-\hat{\vect{d}}\sprod\vect{v}\vprod\hat{\vect{B}}(\vect{r}_\mathrm{A})$
where $\hat{\vect{d}}$ is the atomic electric-dipole operator. Here, the first term is the usual electric-dipole coupling for an atom at rest and the velocity-dependent second term is the so-called R\"ontgen interaction \cite{SYBRef3}. The presence of this leading-order relativistic correction can be understood from the fact the electric field experienced by the moving atom in its own rest frame reads $\hat{\vect{E}}'=\hat{\vect{E}}+\vect{v}\vprod\hat{\vect{B}}$.

In order to induce discriminatory energy shifts, the mirror acting as a model for a general chiral system needs---again according to the Curie symmetry principle---to exhibit chiral properties corresponding to those of the Rydberg system. Such media are a special case of arbitrary linear media, as represented by their frequency-dependent complex-valued nonlocal conductivity tensor $\ten{Q}(\vect{r},\vect{r}',\omega)$, where the quantised electromagnetic fields read \cite{SYBRef4}
\begin{eqnarray}
\label{SYB2}
\hat{\vect{E}}(\vect{r})&=&
\mi\mu_0\int_0^\infty\dif\omega\,
\bigl[\ten{G}\star\hat{\vect{j}}_\mathrm{N}\bigr](\vect{r},\omega)+\operatorname{h.c.},\\
\label{SYB3}
\hat{\vect{B}}(\vect{r})&=&
\mu_0\int_0^\infty\frac{\dif\omega}{\omega}\,\bm{\nabla}\vprod
\bigl[\ten{G}\star\hat{\vect{j}}_\mathrm{N}\bigr](\vect{r},\omega)+\operatorname{h.c.}
\end{eqnarray}
with the magnetic constant $\mu_0$, with $+\operatorname{h.c.}$ implying the addition of the Hermitian conjugate of the previous expression and with $\bigl[\ten{T}\star\vect{v}\bigr](\vect{r})=\int\dif^3s\,
\ten{T}(\vect{r},\vect{s})\sprod\vect{v}(\vect{s})$ and $\bigl[\ten{S}\star\ten{T}\bigr](\vect{r},\vect{r}')=\int\dif^3s\,\ten{S}(\vect{r},\vect{s})\sprod\ten{T}(\vect{s},\vect{r'})$ denoting spatial convolutions for tensor and vector fields.  Here, $\ten{G}$ is the classical Green tensor of rank 2 for the electromagnetic field which satisfies a generalised Helmholtz equation (with $\bm{\delta}$ denoting the Kronecker tensor of rank 2)
\begin{equation}
\label{SYB4}
\biggl[\vect{\nabla}\vprod\vect{\nabla}\vprod-\frac{\omega^2}{c^2}\biggr]\ten{G}
-\mi\mu_0\omega[\ten{Q}\star\ten{G}]=\bm{\delta}
\end{equation}
together with the boundary condition $\ten{G}(\vect{r},\vect{r}',\omega)\to\tenszero$ for $|\vect{r}-\vect{r}'|\to\infty$. It fulfils the completeness relation
\begin{equation}
\label{SYB5}
\mu_0\omega\ten{G}\star\GRe(\ten{Q})\star\ten{G}^\dagger=\GIm(\ten{G})
\end{equation}
where generalised real and imaginary parts of nonsymmetric tensor fields are defined as
$\GRe(\ten{T})=[\ten{T}(\vect{r},\vect{r}')+\ten{T}^\dagger(\vect{r}',\vect{r})]/2$ and
$\GIm(\ten{T})=[\ten{T}(\vect{r},\vect{r}')-\ten{T}^\dagger(\vect{r}',\vect{r})]/(2\mi)$; they reduce to ordinary (component wise) real and imaginary parts for symmetric tensor fields with $\ten{T}^\dagger(\vect{r}',\vect{r})=\ten{T}^\ast(\vect{r},\vect{r}')$. The noise currents $\hat{\vect{j}}_\mathrm{N}$ residing inside the present media can be expressed in terms of bosonic operators $\hat{\vect{f}}$:
$\hat{\vect{j}}_\mathrm{N}=\sqrt{\hbar\omega/\pi}\ten{R}\star\hat{\vect{f}}$
with being an arbitrary solution to $\ten{R}\star\ten{R}^\dagger=\GRe(\ten{Q})$.

With these preparations at hand, we can determine the dispersive energy shift of an atom in a given Rydberg state $|n\rangle$ due to its interaction with the quantum electromagnetic field in an arbitrary environment by following the original approach of Casimir and Polder \cite{SYBRef5}. Assuming the electromagnetic field to be in the vacuum state $|\{0\}\rangle$, we calculate the energy shift
\begin{equation}
\Delta E=\sum_{I\neq\psi}
 \frac{\langle \psi|\hat{H}_{A\mathrm{F}}|I\rangle
 \langle I|\hat{H}_{A\mathrm{F}}|\psi\rangle}
 {E_\psi-E_I}
 \end{equation}
 of the uncoupled state $|\psi\rangle=|n\rangle|\{0\}\rangle$ arising from the above atom--field dispersion interaction within leading, second-order perturbation theory. The relevant intermediate states $|I\rangle=|k\rangle|\vect{1}(\vect{r},\omega)\rangle$ involve single-photon states $|\vect{1}(\vect{r},\omega)\rangle=\vect{f}^\dagger(\vect{r},\omega)|\{0\}\rangle$, so the formal sum consists of a sum over atomic states $|k\rangle$ and integrals over the respective position and frequency arguments. We insert the atom--field coupling into the above formula and retain only the leading-order chiral contributions which are due to cross-terms of electric-dipole and R\"ontgen couplings. Evaluating the matrix elements by means of the field expansions~(\ref{SYB2}) and (\ref{SYB3}) and using the completeness relation~(\ref{SYB5}), we find
\begin{multline}
\label{SYB8}
\hbar\delta_\mathrm{RM}=\frac{\mu_0}{\pi}\sum_{k\neq n}\mathcal{P}\int_0^\infty\dif\omega\,
\frac{\omega}{\omega-\omega_{nk}}\\
\times\operatorname{Re}\bigl[\vect{v}\vprod\vect{d}_{kn}\sprod\bm{\nabla}\vprod
 \ten{G}(\vect{r},\vect{r}_\mathrm{A},\omega)\sprod\vect{d}_{nk}
\\
-\vect{v}\vprod\vect{d}_{nk}\sprod\bm{\nabla}\vprod
 \ten{G}(\vect{r},\vect{r}_\mathrm{A},\omega)\sprod\vect{d}_{kn}\bigr]_{\vect{r}=\vect{r}_\mathrm{A}}
\end{multline}
with atomic transition frequencies $\omega_{nk}=(E_n-E_k)/\hbar$, the electric dipole matrix elements $\vect{d}_{nk}=\langle n|\hat{\vect{d}}|k\rangle$ and the Cauchy principal value $\mathcal{P}$. As a consistency check, we note that the above energy shift hence vanishes when (i) the dipole moments are linearly polarised ($\vect{d}_{nk}=\vect{d}_{kn}$) and hence the atom does not exhibit handedness. Upon splitting the Green tensor $\ten{G}=\ten{G}^{(0)}+\ten{G}^{(1)}$ into its bulk and scattering parts $\ten{G}^{(0)}$ and $\ten{G}^{(1)}$, respectively, the former does not contribute due to the symmetry $\ten{G}^{(0)}(\vect{r},\vect{r}')=\ten{G}^{(0)}(\vect{r}',\vect{r})$. After applying contour-integration techniques and only retaining the dominant resonant contribution from the pole at $\omega=\omega_{nk}$, we find
\begin{multline}
\label{SYB9}
\hbar\delta_\mathrm{RM}=\mu_0\sum_{k<n}\omega_{nk}
\bigl[\vect{v}\vprod\vect{d}_{nk}\sprod\bm{\nabla}\vprod
 \operatorname{Re}\ten{G}^{(1)}(\vect{r}_\mathrm{A},\vect{r}_\mathrm{A},\omega_{nk})\sprod\vect{d}_{kn}\\
 +\operatorname{c.c.}
\end{multline}

Finally, we apply this general result to our specific geometry of a Rydberg atom with a single circularly polarised transition $\vect{d}_{kn}=(d_{kn}/\sqrt{2})(\vect{e}_y+\mi\vect{e}_z)$ rotating in the $yz$ plane travelling with a velocity $\vect{v}=\vect{e}_x$ parallel to the rotation axis at position $\vect{r}_\mathrm{A}=z_\mathrm{A}\vect{e}_z$ above a perfectly reflecting chiral mirror in the $z=0$ plane (see Fig.~\ref{fig:SYB1}). The scattering Green tensor of the mirror is given by \cite{SYBRef6,SYBRef7,SYBRef8}
\begin{multline}
\label{SYB10}
\ten{G}^{(1)}(\vect{r},\vect{r}',\omega)
=\frac{\mi}{8\pi^2}\int\frac{\dif^2k^\parallel}{k^\perp}
\me^{\mi\vect{k}^\parallel\sprod(\vect{r}-\vect{r}')+\mi k^\perp(z+z')}\\
\times \sum_{\sigma,\sigma'=\mathrm{s},\mathrm{p}}r_{\sigma\sigma'}
\vect{e}_{\sigma+}\vect{e}_{\sigma'-}
\end{multline}
where $\vect{e}_{\sigma\pm}$ are unit vectors for incident ($-$) and reflected ($+$)  $\mathrm{s}$- and $\mathrm{p}$-polarised waves and $r_{\sigma\sigma'}$ are the respective reflection coefficients. For a perfect chiral mirror in particular, which rotates the polarisation of incoming light by $\pi/2$ upon reflection, we have $r_\mathrm{ss}=r_\mathrm{pp}=0$, $r_\mathrm{sp}=-r_\mathrm{ps}\equiv r$. Chiral mirrors can be realised by placing a 2d fish-scale \cite{fedotov2006} or split-ring \cite{plum2015} metamaterial at sub-wavelength distance from a metallic mirror, where reflectivities in the range $80\%$--$90\%$ have been reported.
Carrying out the integral over the parallel component of the wave vector $\vect{k}^\parallel$, one easily finds
\begin{equation}
\label{SYB11}
\bm{\nabla}\vprod\ten{G}^{(1)}(\vect{r}_\mathrm{A},\vect{r}_\mathrm{A},\omega_{nk})
=-\frac{\mi c r}{32\pi\omega_{nk}z_\mathrm{A}^3}\,
\begin{pmatrix}1&0&0\\0&1&0\\0&0&2\end{pmatrix}
\end{equation}
in the nonretarded regime $z_\mathrm{A}\ll c/\omega_{kn}$, typically valid for Rydberg atoms. Combining this with the above choices for velocity and dipole moments and fixing the phase upon reflection to be
$r=\me^{\mi\pi/2}=\mi$, we find
$\hbar\delta_\mathrm{RM}=3vd_{kn}^2/(32\pi\varepsilon_0c z_\mathrm{A}^3)$.
Indeed, a moving Rydberg atom with a circular dipole transition exhibits a chiral energy shift near a nonreciprocal mirror and can hence act as a chiral sensor. As required, the detected energy shift changes sign upon replacing the Rydberg system with its opposite enantiomer by reversing either the velocity ($\vect{v}\mapsto -\vect{v}$) or the rotation of the dipole transition ($\vect{d}_{kn}\mapsto\vect{d}_{kn}^\ast$).

To estimate the order of magnitude of the discriminatory chiral energy shift, let us compare it with the ordinary resonant electric energy shift \cite{SYBRef9,SYBRef10}
$\hbar\delta=3d_{kn}^2/(128\pi\varepsilon_0z_\mathrm{A}^3)$ of a stationary Rydberg atom at distance $z_\mathrm{A}$ from a perfectly conducting plate which arises from the first term in the atom--field coupling. The discriminatory term displays the same distance dependence as the standard electric energy shift. Being a leading-order relativistic correction, it is smaller than the latter by a factor $4v/c$ which for a beam with velocity $10^3\,\mathrm{m}/\mathrm{s}$ is of the order $10^{-5}$. Given that electric frequency shifts of Rydberg atoms can be of the order of $10^9\,\mathrm{Hz}$ at a distance of $1\,\mu\mathrm{m}$ \cite{SYBRef11}, we estimate the chiral frequency shifts to be of the order of $10^4\,\mathrm{Hz}$. It can be enhanced by further reducing the distance (where higher-order multipole moments need to be taken into account) or by enhancing the atom--field coupling using microwave cavities. The predicted linear scaling with velocity $v$ is only valid for nonrelativistic speeds. Geometrically, the slope of the helix traced out by the electric dipole vector in Fig.~\ref{fig:SYB1} is $h/(2\pi r)=v/(a_n\omega_{kn})$ with pitch $h=2\pi v/\omega_{kn}$  and radius $r=a_n$ (radius of Rydberg orbital) is very small for Rydberg atoms at moderate speeds, as can be seen by estimating $a_n\approx a_0n^2$ and $\omega_{kn}\approx 2E_\mathrm{R}/(\hbar n^3)$, hence $a_n\omega_{kn}\approx 2\times 10^6(\mathrm{m}/\mathrm{s})/n$.

\paragraph{Conclusion.}
Rydberg atoms are well suited for implementing tailored quantum sensors able to measure small interaction-induced energy shifts. By combining sensitivity with the stability offered by the clock states in the ground state manifold we can implement a sensitive interferometer which should be able to sense motion-induced chirality in achiral atoms. We give estimates of the expected interaction strength by calculating the interaction with a chiral mirror.
\paragraph{Acknowledgements.} We are grateful for discussions with T.~Momose, A.~Salam, F.~Suzuki. This work has been funded by the German Research Foundation (DFG, project number 328961117 -- SFB 1319 ELCH, grant BU 1803/3-1476, S.Y.B.).

\bibliographystyle{apsrev4-1}
\bibliography{references}
\end{document}